\documentclass[12pt]{article}
\title{ On the Current Correlators at Low Temperature}
\author{B.L.Ioffe\thanks{ioffe@itep.ru}\\
 Institute of Theoretical and Experimental Physics,\\ 117218,
Moscow, B.Cheremushkinskaya 25}
\date{}
\begin{document}
\maketitle

\newcommand{\be}{\begin{equation}}
\newcommand{\ee}{\end{equation}}

\def\la{\mathrel{\mathpalette\fun <}}
\def\ga{\mathrel{\mathpalette\fun >}}
\def\fun#1#2{\lower3.6pt\vbox{\baselineskip0pt\lineskip.9pt
\ialign{$\mathsurround=0pt#1\hfil##\hfil$\crcr#2\crcr\sim\crcr}}}

\vspace{1cm}

\begin{abstract}
It is demonstrated, that the calculations of current correlators
at low $T$, performed in Ref.'s \cite{1},\cite{2},\cite{3a} by the
methods of current algebra, are correct contrary to the statements
of the recent paper by Mallik \cite{3}.
\end{abstract}

The correlators of hadronic currents were considered in
Ref.\cite{1,2} at finite and small temperature $T$ in order of
$T^2$ (see also \cite{3a}). The method of consideration was the
partial conservation of the axial current (PCAC) and current
algebra. The main result obtained was the following. If the
correlator is represented as the sum over intermediate hadronic
states (resonances), then in order $T^2$ there is no shifts of the
resonances masses (poles of the correlators) and the only
interesting physical phenomenon which occurs in this order is the
parity mixing, i.e. the admixture of states with opposite parity
in the given channel and, in some cases, also an isospin mixing.
The thermal correlators of vector and axial isovector currents
were considered: \be C^J_{\mu\nu} (q,T) = i\int d^4xe^{iqx}\sum_n
\langle n\mid T\{ J^a_{\mu}(0),~J^a_{\nu}(x)\} \exp [~(\Omega
-H)/T~]n\rangle\label{1}\ee ($J=V,A;a$ is isospin index). Here
$e^{-\Omega}= \sum_n \langle n \mid \exp (-H/T)\mid n\rangle$ and
the sum is over  the full set of the eigenstates  of the
Hamiltonian $H$. It was demonstrated that in the limit of massless
pions in order of $T^2$ $(-q^2=Q^2 \gg T^2$) \be C^V_{\mu\nu}
(q,T) =(1-\varepsilon)C^V_{\mu\nu} (q,0) +\varepsilon
C^A_{\mu\nu}(q,0)\label{2}\ee \be C^A_{\mu\nu}(q,T) =
(1-\varepsilon) C^A_{\mu \nu} (q,0) +\varepsilon C^V_{\mu\nu}
(q,0),\label{3}\ee where $\varepsilon = T^2/6
F^2_{\pi},~F_{\pi}=93$ MeV is the pion decay constant. The
derivation of (\ref{2}),(\ref{3}) was based on the fact, that in
order $T^2$ it is sufficient to account for contributions of two
lowest states in (\ref{1}) -- vacuum and one pion. Then the factor
$T^2$ arises simply from  the one -- pion phase space. The
contributions  of higher excited states are suppressed as
$T^k,~k=4,6...$ (in case of pions) or $e^{-m_h/T}$, where $m_h$
are the masses of massive hadronic states. The mentioned above
result arises immediately from (\ref{2}),(\ref{3}) after
decomposition of $C^J_{\mu\nu}(q,T)$ and $C^J_{\mu\nu}(q,0)$ into
the sum over intermediate states.

In the paper by Mallik \cite{3} it was found that the  case of
degenerate (or nearly degenerate) intermediate states requires
special consideration. (Such case was not considered in
\cite{1,2,3a}). In \cite{3} the correlator of the vector currents
was considered, $\rho$ and $\omega$-mesons are assumed to be
degenerate, $m_{\rho}=m_{\omega}$, their widths were neglected. It
was found that $\rho$ and $\omega$ mass shifts are absent in
agreement with \cite{1,2}, but the residue at the $\rho$ meson
pole in order $T^2$ changes its universal form $1- \varepsilon$ to
$1 - (1 + g^2_1/3) \varepsilon$, where $g_1$ is the $\rho \omega
\pi$ coupling constant. It is important to mention that this
result is valid only at $q^2 = m^2_{\rho}$. If $\vert q^2 -
m^2_{\rho}\vert$ is large enough, $\vert q^2 - m^2_{\rho} \vert
\gg m_{\rho} T$, then the proportional to $g^2_1$ term in the
residue is multiplied by the factor of order $T^2/(q^2 -
m^2_{\rho})$, its contribution is of order $T^4$, i.e. of the
order of the neglected terms. Therefore, the domain in $q^2$,
where the proportional to $g^2_1$ correction is essential, is
narrow at small $T$. For example, if we would like to apply the
procedure, common in the vector dominance model, and extrapolate
the correlator  from $q^2 = m^2_{\rho}$ to $q^2 = 0$, we shall put
$g^2_1 = 0$. The account of the $\rho$-meson width drastically
changes the situation. In this real physical situation $\rho$ and
$\omega$-mesons are no more degenerate, $\rho \to \omega \pi$
transition proceeds through the emission  of pions in the
$p$-wave, the small numerator which arises from the $\rho \to
\omega \pi \to \rho$ transition is not compensated by the
denominator. The effect of $\rho \to \omega\pi \to \rho$
transitions results in appearance of the $T^4$ terms in $C^V_{\mu
\nu} (q,T)$ and the formulae (\ref{2}), (\ref{3}) are intact. So,
the result [Eq.(1.1)] of the Ref.\cite{3} is very limited -- only
in a model, where $\rho$ width is neglected and even there in the
narrow interval of $q^2$.

Turn now to the second problem, discussed in \cite{3}: the $T^2$
correction to nucleon coupling $\lambda$ in the correlator of
nucleon currents \cite{4}-\cite{6a}. This correction was
calculated in Ref.\cite{5} as \be \lambda^T=\lambda \left \{1 -
\frac{(g^2_A+1)}{32} \frac{T^2}{F^2_{\pi}}\right\},\label{4}\ee
where $g_A$ is the nucleon axial coupling constant. In
Ref.\cite{3} it is claimed, that $g^2_A$ term in (\ref{4}) would
be missed in the calculations performed by the method of
Ref.\cite{1,2}. This is, however, not true. The correlators of
baryon currents were considered in \cite{2}, but only the mixing
of correlators with different isospin and opposite parity was
discussed. No calculation of the residue at nucleon pole was
performed. If such calculation  would be done by PCAC technique,
the proportional to $g^2_A$ term would arise from \be
\frac{1}{F^2_{\pi}} \int d^4 x e^{iqx}  \int d^4 y e^{iky} \int
d^4ze^{ikz} k_{\mu} k_{\nu} \langle 0\mid T\{\eta(x),
j^A_{\mu}(y), j^A_{\nu}(z), ~\bar{\eta}(0)\} \mid 0\rangle,
\label{5}\ee where $j^A_{\mu}$ and $\eta$ are axial and nucleon
currents, $k$ is the pion momentum. Usually in PCAC the term
(\ref{5}) is omitted, because it is higher order in $k$ in
comparison with the term with equal-time commutator, arising from
differentiation of $\theta$-function. But in the case under
consideration  $j^A_{\mu}$ has a pole in momentum space,
corresponding to one-pion state, $j_{\mu}\sim k_{\mu}/k^2$, the
term (\ref{5}) is of zero order in $k$ and must be accounted.
Because of the Goldberger-Treiman theorem the contribution of
(\ref{5}) to correlator of nucleon currents is proportional to
$g^2_A$. Therefore, the critical remark in this part of
Ref.\cite{3} is not correct.

The support is acknowledged from CRDF grant RUP2-2621-MO-04 and
RFBR grant 06-02-16905.

 \end{document}